# The measured speed in the evanescent regime reflects the spatial decay of the wavefunction, not particle motion

Weixiang Ye

Center for Theoretical Physics, School of Physics and Optoelectronic Engineering, Hainan University, Haikou 570228, China

e-mail: wxy@hainanu.edu.cn

**Arising from:** V. Sharoglazova *et al. Nature* **643**, 67–72 (2025).

The recent paper by Sharoglazova et al. [1] reports an energy-dependent parameter $v$ extracted from the spatial distribution of photons in a coupled-waveguide experiment. The authors interpret $v$ as the "speed" of quantum particles, even in the classically forbidden regime, and claim that its finite value contradicts the Bohmian-mechanics prediction [2] of zero particle velocity. This challenge arises from a fundamental conflation of two distinct concepts: a geometric property of the wavefunction and the kinematical velocity of a point-like particle. We demonstrate that $v$ quantifies the spatial decay rate of the wave-field amplitude, a property of the guiding field, not the velocity of Bohmian particles. The experiment therefore does not challenge but rather illustrates the clean ontological separation between the wave and particle aspects inherent to Bohmian mechanics.

The experiment yields two distinct results: a parameter $v$ from intensity-based population transfer, and a phase-gradient velocity $v_S$ from interferometry. For evanescent states ($\Delta < 0$), they find $v_S \approx 0$ while $v > 0$. The claimed contradiction rests on identifying $v$ as the particle's kinematical speed. This identification is incorrect within Bohmian mechanics, where particle motion is uniquely defined by the guiding equation. The measurement protocol, recording only scalar intensities $|\psi_m(x)|^2$ and $|\psi_a(x)|^2$, is operationally blind to the phase gradient $\nabla S$ that governs particle dynamics. This separation is formalized via the weak value of momentum [3]. For a wavefunction in polar form $\psi = R e^{iS/\hbar}$, the weak value is

$$p_w(x) = \nabla S(x) - i\hbar \frac{\nabla R(x)}{R(x)}. \qquad (1)$$

Its real and imaginary parts are independently accessible:

$$v_S(x) = \frac{1}{m}\mathrm{Re}[p_w] = \frac{\nabla S}{m}, \quad \frac{\hbar\kappa}{m} = \frac{1}{m}|\mathrm{Im}[p_w]|, \kappa \equiv \left|\frac{\nabla R}{R}\right|. \qquad (2)$$

Interferometry measures $v_S$, which equals the Bohmian particle velocity $v_B$ [2,4]. The population-transfer measurement probes $|\mathrm{Im}[p_w]|/m$. The model in Ref. [1] gives initial growth $\rho_a(x) \approx (J_0 x/v)^2$, which is equivalent to $\rho_a(x) \approx (\kappa_1 x)^2$ with $\kappa_1 = mJ_0/(\hbar\kappa)$, leading to

$$v = \frac{J_0}{\kappa_1} = \frac{\hbar \kappa}{m}. \qquad (3)$$

Equation (3) explicitly defines $v$ as a measure of the wavefunction's spatial decay rate $\kappa$. The amplitude decay constant for $\Delta < 0$ is determined by the Schrödinger equation for the evanescent tail: $\hbar^2 \kappa^2/(2m) = |\Delta|$. Substituting this into Eq. (3) yields $v = \sqrt{2|\Delta|/m}$. Therefore, the observed relation for $|\Delta| \gg \hbar J_0$ is a direct consequence of wave dynamics, which Bohmian mechanics fully embraces. For $\Delta < 0$, $v_B = 0$ correctly indicates stationary particles, while $v = \hbar \kappa / m > 0$ quantifies the spatial decay of the guiding wave-field.

One might argue that the steady-state distribution encodes a history of motion. In Bohmian mechanics, its establishment involves a small fraction of particles temporarily entering the forbidden region before turning back [5, 6]. Crucially, the experimental protocol for extracting $v$—fitting the steady-state spatial profile $\rho_a(x) \propto x^2$—is insensitive to the time-dependent details of how this profile was established. The parameter $v$ is a property of the stationary wavefunction's spatial form, not a measure of the transient kinematics of the scattering particles. The operational definition of a particle's velocity in Bohmian mechanics is given by $v_B = j/\rho$, derived from the continuity equation and the ontological commitment to point particles with continuous trajectories. Any other quantity derived from a different operational procedure, such as fitting a spatial profile, cannot be logically identified with this velocity without altering the theory's foundational structure. The supposed "challenge" therefore does not stem from a conflict between theory and experiment, but from mistakenly attributing an operational measure of wave-field geometry to the ontologically distinct concept of particle motion. Any theoretical modification that forces particle velocities to match this geometric parameter would sacrifice the internal consistency and explanatory power of Bohmian mechanics.

This clarifies the dwell time. The experimental timescale $\tau_\lambda = \lambda/v$ (with $\lambda = 1/\kappa$) is a geometric construct: $\tau_\lambda = m/(\hbar \kappa^2)$. In Bohmian mechanics, the dwell time $\tau_{\text{dwell}} = N/j_{\text{in}}$ diverges because the probability current $j_{\text{in}} = 0$ for a stationary scattering state [6]. The agreement between data and $\tau_\lambda$ confirms the geometric relation (3), not particle motion. The numerical coincidence arises because the standard quantum-mechanical calculation of $\tau_{\text{dwell}}$, when performed using a specific (and debatable) decomposition of the probability current, yields an expression mathematically similar to $\tau_\lambda$. This does not elevate $v$ to a particle velocity.

The detection of signal in the classically forbidden region ($x > 0$) is an ensemble effect. The steady-state intensity $I(x) \propto |\psi(x)|^2$ is built from many independent events. This distribution is operationally inferred from the spatial pattern of photon detections at the camera, which reflects the time-integrated intensity converging to the stationary $|\psi(x)|^2$. In each run, a Bohmian particle's trajectory is confined to $x \leq 0$ in the steady state.

Proposed modifications to the guiding equation [7, 8] attempt to force Bohmian trajectories to match the geometric parameter $v$, thereby conflating wavefunction geometry with particle kinematics. Such adjustments abandon the principled ontological separation that is a virtue of the standard theory. Crucially, they accept the very conflation that constitutes the misinterpretation of the experiment. Within the standard Bohmian framework, the velocity field $v_B$ is uniquely determined by the requirement of consistency with the Schrödinger equation's continuity equation and the existence of unique particle trajectories. Any ad hoc modification to make $v_B$ non-zero in the evanescent region would break this consistency or introduce arbitrary, untestable assumptions about particle dynamics.

The experiment's real significance is that it operationally distinguishes two quantities: the phase-gradient velocity $v_S$ (particle motion) and the amplitude-decay parameter $v$ (wave-field geometry). In standard quantum mechanics, this distinction is only implicit in the formalism. In Bohmian mechanics, however, it is directly reflected in the clear ontology of point particles guided by a wave. The theory naturally accommodates a stationary particle distribution ( $v_B = 0$ ) together with an exponentially decaying wave field ( $v > 0$ ). The claimed "non-uniqueness" of the guiding equation is not a liability but a mathematical red herring: any alternative flow field that artificially equates the geometric $v$ with particle velocity would violate fundamental principles like quantum equilibrium and Galilean covariance, and would thus constitute a different, inferior theory. Far from challenging Bohmian mechanics, the experiment vividly demonstrates its capacity to parse quantum phenomena into distinct yet coherent particle and wave aspects—a clarity that is precisely the goal of a realist interpretation.

In summary, the experiment by Sharoglazova et al. is a sophisticated measurement of wavefunction geometry, misinterpreted as a measurement of particle kinematics. This misidentification undermines the claimed challenge. Within Bohmian mechanics, the clean separation of $v_S$ (particle velocity) from $v$ (wave-field geometric rate) validates its ontological structure [9]. The energy dependence of $v$ is a prediction of the wave dynamics it embraces. This experiment strikingly demonstrates the Bohmian framework, where $v$ finds its natural meaning as the magnitude of the momentum weak value's imaginary part—a weak actual value characterizing the wave-field's local property [10].

**References (Main Text)**

**Supplementary Information**

**SI 1: Derivation of the wave-geometric speed and operational separation**

The stationary equations for the coupled-waveguide system are [R1]:

$$E\psi_m = -\frac{\hbar^2}{2m}\frac{d^2\psi_m}{dx^2} + V_0\psi_m + \hbar J_0(\psi_a - \psi_m),$$
$$E\psi_a = -\frac{\hbar^2}{2m}\frac{d^2\psi_a}{dx^2} + V_0\psi_a + \hbar J_0(\psi_m - \psi_a),$$
(S1)

with $\Delta = E - V_0 + \hbar J_0$.

Following the solution in Ref. R1, the parameter $v_J$ extracted from the initial population growth $\rho_a(x) \approx (J_0 x / v_J)^2$ is

$$v_J = \sqrt{\frac{|\Delta \pm \sqrt{\Delta^2 - (\hbar J_0)^2}|}{m}},$$
(S2)

where the sign is chosen for continuity. In the weak-coupling limit ($|\Delta| \gg \hbar J_0$), this reduces to

$$v \approx \sqrt{2|\Delta|/m}.$$
(S3)

The amplitude decay constant $\kappa$ for the component $\psi_m$ is $\kappa = \sqrt{2m|\Delta|}/\hbar$ in the same limit. Comparing with Eq. (S3) establishes the identity used in the main text:

$$v \approx \frac{\hbar \kappa}{m}.$$
(S4)

This confirms that the experimentally fitted parameter $v$ is fundamentally a measure of the wavefunction's spatial structure, related to its amplitude decay rate $\kappa$.

**SI 2: On the uniqueness of Bohmian velocity and the nature of the steady state**

In Bohmian mechanics, the particle velocity is not an arbitrary quantity but is uniquely fixed by the conjunction of the Schrödinger equation and the ontology of point particles. The continuity equation, derived from the Schrödinger equation, is:

$$\frac{\partial |\psi|^2}{\partial t} + \nabla \cdot \mathbf{j} = 0, \mathbf{j} = \frac{\hbar}{m}\text{Im}(\psi^*\nabla\psi).$$
(S5)

For a system of point particles with well-defined positions $\mathbf{x}(t)$, the probability density $|\psi|^2$ is interpreted as the density of particles. The natural and unique identification that makes the particle dynamics consistent with the evolution of $|\psi|^2$ is to set the velocity field $\mathbf{v}_B$ equal to the probability current divided by the density:

$$\mathbf{v}_B(\mathbf{x}, t) = \frac{\mathbf{j}(\mathbf{x}, t)}{|\psi(\mathbf{x}, t)|^2}.$$
(S6)

Any other choice would lead to a mismatch between the flow of particles and the flow of probability, violating the statistical predictions of quantum mechanics (quantum equilibrium). This derivation is independent of whether the wavefunction is stationary or time-dependent.

The experiment operates in a quasi-steady-state regime. The reported 26 ns pulse duration is much longer than the characteristic system timescales (e.g., the photon lifetime of 270 ps and the scattering transient time, estimated to be on the order of picoseconds). Therefore, the measured spatial intensity profile is indistinguishable from that of a true stationary solution of the Schrödinger equation. The fitting procedure used to extract $v$ acts on this spatial profile alone. It is mathematically decoupled from the time-dependent process that established this profile. Consequently, the extracted parameter $v$ is a functional of the stationary wavefunction's shape, not a time-averaged kinematic quantity. Even if one considers the transient dynamics where Bohmian trajectories briefly enter the forbidden region, the final fitted value of $v$ depends only on the asymptotic stationary state they settle into, not on the details of those transient trajectories. Any minor differences between the time-averaged density of a wave packet and the ideal stationary density are of higher order in the small parameter $\sigma_x \kappa$ (where $\sigma_x$ is the wave-packet width), and thus negligible for the extraction of $\kappa$ from the initial quadratic growth region near $x = 0$.

**SI 3: Detailed analysis of dwell time and the conflation of quantities**

The dwell time $\tau_{\text{dwell}}$ is formally defined in scattering theory as $\tau_{\text{dwell}} = N/j_{\text{in}}$, where $N$ is the integrated probability density in the barrier region and $j_{\text{in}}$ is the incident probability current.

- **In Bohmian mechanics:** For a stationary, real wavefunction (as in the evanescent regime), $\mathbf{j} = 0$ everywhere. Thus, $j_{\text{in}} = 0$, leading to $\tau_{\text{dwell}} \to \infty$. This is a direct consequence of the guiding equation and the stationary-state condition.
- **In the standard scattering calculation:** One often decomposes the standing wave in front of the barrier into incoming and outgoing parts. The incident current is then calculated as $j_{\text{in}} = (\hbar k/m) \mid A_{\text{in}} \mid^2$, where $A_{\text{in}}$ is the amplitude of the incoming wave component. For a reflection coefficient $R = 1$, one finds $j_{\text{in}} = (\hbar k/m)$. Using this, one obtains a finite dwell time: $\tau_{\text{dwell}}^{\text{QM}} = \frac{\int |\psi|^2 dx}{j_{\text{in}}} = \frac{m}{\hbar k \kappa}$.
- **The experimental construct:** The experiment calculates a time $\tau_\lambda = \lambda/v = (1/\kappa)/(\hbar\kappa/m) = m/(\hbar\kappa^2)$. Noting that for a high barrier, $\kappa \approx k$, we see $\tau_\lambda \approx \tau_{\text{dwell}}^{\text{QM}}$.

The key point is that $\tau_\lambda$ is constructed from two geometric parameters, $\lambda$ and $v$, both derived from the wavefunction's spatial shape. Its numerical agreement with $\tau_{\text{dwell}}^{\text{QM}}$ is a mathematical consequence of the Schrödinger equation's structure (specifically, the relation $\hbar^2\kappa^2/(2m) = \mid \Delta \mid$). This agreement does not imply that $v$ is the particle speed, any more than it implies that $\lambda$ is the particle's turning point. It simply reflects an internal consistency within the wave dynamics.

**SI 4: Critique of modified guiding equations and bidirectional models**

Proposals to modify the guiding equation to yield non-zero velocities in evanescent regions [R2, R3] face severe conceptual difficulties within the Bohmian framework.

1. **Violation of Quantum Equilibrium:** The standard guiding equation (S6) is uniquely selected by the requirement that an ensemble of particles initially distributed as $|\psi|^2$ remains so distributed (equivariance). Most ad hoc modifications break this property, requiring an inexplicable, non-equilibrium initial distribution to match quantum predictions.
2. **Arbitrariness:** The continuity equation $\partial_t \rho + \nabla \cdot (\rho \mathbf{v}) = 0$ determines only the divergence of the current. The standard choice $\mathbf{v} = \mathbf{j}/\rho$ is the simplest and most natural that also makes the dynamics Galilean and time-reversal covariant. Alternative choices, such as adding a divergence-free flow, introduce arbitrary functions not determined by the wavefunction, rendering the theory non-predictive.
3. **Galilean Covariance:** The standard Bohmian dynamics is explicitly Galilean covariant [R4]. The velocity field $v_B = j/\rho$ transforms correctly under changes of inertial frame. Any replacement of $v_B$ with a term proportional to $\nabla R/R$ (which is not a covariant four-vector under Galilean boosts) would destroy this fundamental symmetry, a fatal flaw for any candidate fundamental theory.
4. **Misidentification of Ontology:** Models like the "bidirectional" model [R3] effectively reinterpret the particle ontology. They postulate particles that switch direction or possess internal states to mimic the wave's spatial decay profile. This goes beyond guiding a point particle with a wavefunction; it constructs a complicated pseudo-particle dynamic whose sole purpose is to match the geometric parameter $v$. This abandons the original simplicity and explanatory power of Bohmian mechanics.

The standard Bohmian interpretation provides a clear explanation: the wavefunction has an exponentially decaying tail in the forbidden region (a *wave* property), which influences the probability density for particle positions. Particles themselves do not move in this region in the steady state. The fact that the wavefield's spatial decay rate ($v$) is energy-dependent is a feature of the wave dynamics, fully accounted for by the Schrödinger equation. There is no need, and no principled reason within the theory, to attribute this decay rate to particle motion.

**SI 5: The weak actual value and the interpretation of $v$**

The concept of *weak actual values* within Bohmian mechanics [R5] provides a consistent interpretive framework for the measured parameter $v$. For an observable $\hat{A}$, its weak actual value along a Bohmian trajectory $Q(t)$ is defined as:

$$a_w(t) = \frac{\text{Re}[\psi^*(Q(t),t)(\hat{A}\psi)(Q(t),t)]}{|\psi(Q(t),t)|^2}. \quad (S7)$$

Its ensemble average under quantum equilibrium equals $\langle \hat{A} \rangle$. For the momentum operator $\hat{p} = -i\hbar\nabla$, the weak actual value is:

$$p_w(t) = \nabla S(Q(t), t) - i\hbar \frac{\nabla R(Q(t), t)}{R(Q(t), t)}. \quad (S8)$$

The real part, $\nabla S$, is the Bohmian particle momentum. The magnitude of the imaginary part, $\hbar |\nabla R/R| = \hbar\kappa$, is precisely the quantity measured as $v$ (up to a factor of $1/m$). This formalism cleanly separates the two contributions: the *particle momentum* (real, directional) and the *wave momentum* (imaginary, related to spatial decay). The experiment directly measures a property of the latter. Therefore, the experiment is not measuring the kinematical velocity of Bohmian particles, but rather a specific attribute of their guiding wavefield. This striking operational separation validates, rather than challenges, the Bohmian framework's capacity to distinguish between the dynamics of the particle and the geometry of its associated wave.

**References (Supplementary Information)**